\documentclass[usenatbib]{mn2e}
\usepackage{amssymb}

\usepackage{graphicx}
 
\title[On the mass of the neutron star in Cyg X-2]
{On the Mass of the neutron star in Cyg X-2}

\author[J. Casares et al.]
{J. Casares$^1$, J.I. Gonz\'alez Hern\'andez$^{2,3}$, G. Israelian$^1$, R.
Rebolo$^{1,4}$\\
$^1$ Instituto de Astrof\'{\i}sica de Canarias, E-38200 La Laguna, Tenerife, 
Spain\\
$^2$ Cosmological Impact of the First STars (CIFIST) Marie Curie Excellence
Team\\ 
$^3$ GEPI, Observatoire de Paris, CNRS, Universit\'e Paris Diderot; Place
Jules Janssen 92190, France\\
$^4$ Consejo Superior de Investigaciones Cient\'\i ficas, Spain}

\begin{document}

\maketitle

\begin{abstract} 
\noindent We present new high resolution spectroscopy of the low mass X-ray 
binary Cyg X-2 which enables us to refine the orbital solution and rotational
broadening of the donor star. In contrast with Elebert et al (2009) we find a 
good agreement with results reported in Casares et al. (1998). 
We measure $P=9.84450\pm0.00019$ day, $K_2=86.5\pm1.2$ km s$^{-1}$ and 
$V \sin i=33.7\pm0.9$ km s$^{-1}$. 
These values imply $q=M_{2}/M_{1}=0.34 \pm 0.02$ and $M_{1}=1.71\pm 0.21$ 
M$_{\odot}$ (for $i=62.5 \pm 4^{\circ}$). Therefore, the neutron star in 
Cyg X-2 can be more massive than canonical. We also find no evidence 
for irradiation effects in our radial velocity curve which could explain the 
discrepancy between Elebert et al's and our $K_2$ values.

\end{abstract}


\section{Introduction}

Dynamical studies in low mass X-ray binaries (LMXBs hereafter) 
offer a promissing route 
to test the equation of state of nuclear matter. Soft equations of state are not
stable above 1.6 M$_{\odot}$ \citep[e.g.][]{brown94} and hence finding a neutron 
star more massive than this limit would be a major advance in our knowledge of 
nuclear matter physics. LMXBs and, in particular, accreting millisecond pulsars 
(AMPs) are expected to harbour the most massive neutron stars because of 
the sustained accretion of matter during their long 
lifes \citep{vandenheuvel95}. 
Unfortunately, dynamical studies are hampered by (1) the
overwhelming accretion luminosity in persistent LMXBs (which swamps the donor 
star's spectrum) and (2) the extreme faintness of the 
companion star in transient AMPs during quiescence
\citep{davanzo09}. 

A promissing route to overcome these limitations was proposed by 
\cite{steeghs02}. X-ray irradiated donor stars can be betrayed by the detection 
of high excitation fluorescence NIII and CIII lines within the Bowen blend at 
$\simeq$4634-50 \AA . The Doppler shift of these narrow lines traces the orbit of
the donor star and, therefore, can yield dynamical constraints in  
persistent LMXBs and transient AMPs during outburst. As a result of this strategy some good candidates for 
massive neutron stars have been found, namely X1822-371 \citep{munoz05} and 
Aql X-1 \citep{cornelisse07}.   

In a few long period ($P>$ 1 day) systems, the donor is an 
evolved star which
can be spectroscopically detected over the irradiated accretion disc. 
Cyg X-2 represents one of this exceptional cases.
Its first orbital solution was presented by  
\cite{cowley79} who show that the donor star is an A5-F2 III orbiting 
the neutron star in 9.843 days with a projected velocity $K_2=87\pm3$ km 
s$^{-1}$. Almost 20 years later, this was refined by \cite{casares98} 
(C98) 
who find 
an A9 donor with orbital parameters $P_{\rm orb}=9.8444 \pm 0.0003$ d and 
$K_2=88.0\pm1.4$ km s$^{-1}$. This work also reports the first determination of
the rotational broadening of the donor's absorption features ($V \sin i = 34.2 
\pm 2.5$ km s$^{-1}$) which, in turn, implies a binary mass ratio 
$q=M_{2}/M_{1}=0.34 \pm 0.04$. This value
is remarkable as it implies a peculiar donor star, very undermassive for
its spectral type and the probable outcome of an intermediate mass binary 
\citep{king99, podsiadlowski00, kolb00}. But the implications for the 
compact object's mass are also remarkable. The orbital solution, combined with 
inclination constraints $i=62.5\pm4^{\circ}$ derived through ellipsoidal fits 
to UBV light curves gives a neutron star mass of 1.78 $\pm$ 0.23 M$_{\odot}$ 
\citep{orosz99} and, hence, a good candidate for a massive neutron star.  
This result has been recently challenged by \cite{elebert09} 
(E09) 
who measure 
$K_2=79\pm3$ km s$^{-1}$ and hence $M_1$=1.5 $\pm$ 0.3 M$_{\odot}$.

Here we present new high resolution spectroscopic observations of Cyg X-2 
with the main aim of refining the rotational broadening of the donor star and
update the orbital parameters. In contrast with E09, we find a
good agreement with previous results reported in C98 which give
support to the presence of a massive neutron star in Cyg X-2.   

\section{Observation and Data Reduction}

Cyg X-2 was observed on the nights of 25-26 July 1999 using the Utrech Echelle 
Spectrograph (UES) attached to the 4.2~m 
William Herschel Telescope (WHT) at the Observatorio del Roque de Los Muchachos. 
Ten 1800-3600s spectra were obtained with the E31 echelle grating and 2Kx2K 
SITe1 detector, covering the spectral range $\lambda\lambda$5300-9000 \AA . 
We selected a 1" slit and a factor 2 binning in the spectral direction,
resulting in 10 km s$^{-1}$ resolution. ThAr arc images
were observed each night for the purpose of wavelength calibration. The final
wavelength calibration was verified using the OI skylines at $\lambda$5577.34
and $\lambda$6300.304 and variable velocity offsets were found between 0.2-5 
km s$^{-1}$. These were corrected from every individual spectrum. 

Eleven additional spectra were obtained on the nights of 31 July 1999 and 9 
July 2000 with the ISIS double-arm spectrograph on the WHT. Here we employed 
the 1200B grating on the blue arm and different central wavelengths resulting 
in wavelength coverages within $\lambda\lambda$3550-6665. 
A 1" slit was selected yielding spectral resolutions in the
range 34-54 km s$^{-1}$. 
Frequent comparison CuAr+CuNe arc lamp images were taken every night for
wavelength calibration. This was tested against sky lines and it was found 
to be accurate to within 4 km s$^{-1}$. These small offsets were nevertheless 
corrected from the individual spectra.  
The ISIS red arm was always centered redwards of 7600 \AA~ but is not
used in this paper due to its lower spectral resolution and the paucity 
of absorption lines, which result in larger errors in the cross-correlation 
analysis. The red arm spectra were mainly taken for the sake of abundance
analysis and will be reported elsewhere (Gonz\'alez Hern\'andez et al. 2009). 
A log of the observations is presented in Table \ref{tabobs}. 

For the purpose of radial velocity analysis we observed the stellar template 
HR 114 (A7 III) using all different instrumental configurations. 
This star has an intrinsic $V \sin i$ of 21 km s$^{-1}$ i.e. comparable to
the rotational broadening of the donor star in Cyg X-2 (C98).
Therefore, to refine our previous rotational broadening  we  decided also to 
observe the F3 V star HR 6189 with UES, which has a reported upper limit 
$V \sin i< 15$ km s$^{-1}$.  

All the images were processed following standard debiasing and flat-fielding, 
and the spectra subsequently extracted using conventional optimal extraction 
techniques in order to optimize the signal-to-noise ratio of the output 
\citep{Horne86}.  

\section{Revisiting the Rotational Broadening}

In C98 we measured the rotational broadening of the donor star's
features in Cyg X-2  using 25 km s$^{-1}$ resolution spectra and found 
$V \sin i=34.2 \pm 2.5$ km s$^{-1}$. Here we revisit this determination using 
the 10 km s$^{-1}$  UES spectra. We broadened the template star HR 6189 from 5 
to 50 km s$^{-1}$ in steps of 1 km s$^{-1}$, using a Gray profile \citep{Gray92} 
and a limb-darkening coefficient $\epsilon=0.5$ appropriate for our spectral 
type and wavelength range. The broadened templates were rectified to the
continuum (usig a low-order spline fit) and multiplied by factors 
$f<1$, to account for their fractional contribution to the total light. These 
were subsequently subtracted from the Doppler corrected average of Cyg X-2, 
obtained using the orbital solution given in Sec. 4. The Cyg X-2 average was 
also rectified through fitting a low order spline to the continuum, after
masking out the main emission and atmospheric/IS absorption lines. 
The optimal broadening, 
based on a $\chi^{2}$ test on the residuals, is found for $34.6 \pm 0.1$ 
km s$^{-1}$. A potential source of 
systematics is the assumption of continuum limb-darkening coefficient in the
computation of the rotational profile. Absorption lines in late-type
stars are expected to have smaller core limb-darkening coefficients than the
continuum \citep{collins95} and, therefore, assuming the continuum
limb-darkening coefficient could bias the result. This was tested by
repeating the same analysis using zero limb-darkening as a conservative lower
limit and obtain $V \sin i=32.8\pm0.1$ km s$^{-1}$. Therefore, we decide to
adopt the mean of the two limb-darkening values as a safe estimate of the true 
rotational broadening in Cyg X-2 i.e. $V \sin i=33.7\pm0.9$ km s$^{-1}$. This 
is in excellent agreement with our determination in C98. 

\begin{figure}
\centering
\includegraphics[width=68mm,angle=-90]{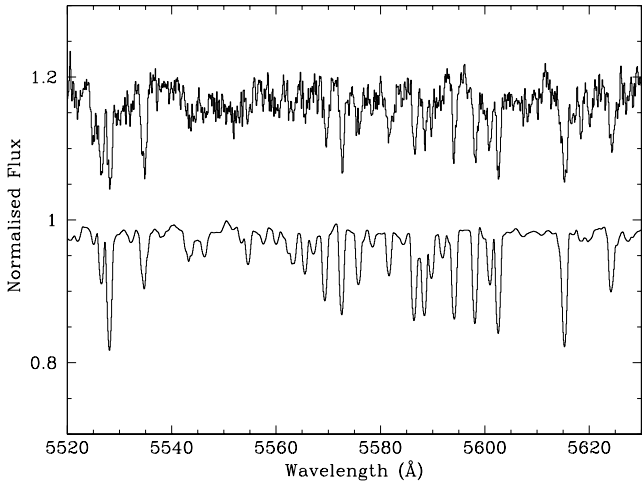}
\caption[]{A section of the Doppler corrected average of the Cyg X-2 UES 
spectra in the rest frame of the donor star (top), together with the F3V 
template HR 6189 broadened by 33.7 km s$^{-1}$ (bottom).}
\label{figopts}
\end{figure}

\section{Updated Orbital Solution and Masses}

\begin{figure}
\centering
\includegraphics[width=68mm,angle=0]{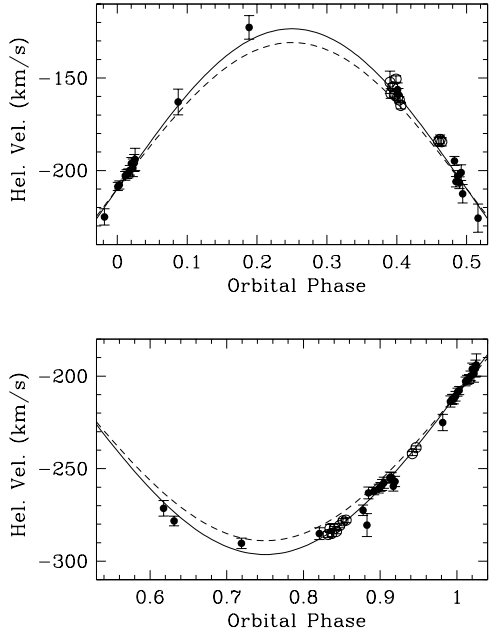}
\caption{Radial velocity curve folded on the ephemeris listed in Sect. 4 with
the best sine-wave solution overplotted. New UES and ISIS velocities from 
1999-2000 are marked with open circles whereas
solid circles indicate velocities from the 1993-1997 ISIS database. The dashed
line depicts the orbital solution of E09.}
\label{figrv}
\end{figure}

We rectified the 21 individual spectra by subtracting a low-order spline fit 
to the continuum, after masking out the main emission and absorption features. 
The UES spectra were subsequently rebinned into a uniform velocity 
scale of 5 km s$^{-1}$ pix$^{-1}$ and the ISIS spectra into 27 km 
s$^{-1}$ pix$^{-1}$. 
Radial velocities were extracted through cross-correlation of every 
individual spectrum of Cyg X-2 with its corresponding A7III template HR 114, 
observed with identical instrumental setup. The template spectra were 
previously broadened to 33 km s$^{-1}$ to match the width of the donor 
photospheric lines (see previous section). Cross-correlation was performed in 
the spectral regions free from emission and telluric absorption features.
For consistency, new radial velocities were extracted from the old 
database (C98) using a contemporaneous 
observation of the same template HR 114, obtained for the purpose of spectral 
classification. Radial velocities were extracted following the 
method of \cite{ton79}, where parabolic fits were performed to the peak of the 
cross-correlation functions, and the uncertainties are purely statistical. 
The final errorbars were multiplied by a factor 2 so that the 
minimum reduced $\chi^{2}$ of a sinewave fit model is 1.0. 
This yields the following parameters 

$$ \gamma= -209.9 \pm 1.4~ {\rm km s}^{-1}$$
$$ P= 9.84450\pm 0.00019~ {\rm d}$$
$$ T_{0} = 2451387.148 \pm 0.018$$
$$ K_{2} = 86.5 \pm 1.2~ {\rm km s}^{-1}$$

\noindent
where $T_{0}$ corresponds to the inferior conjunction of the donor star.
All quoted errors are 68 per cent confidence. The systemic velocity $\gamma$ 
has been corrected from the radial velocity of HR 114, that we take as 
-10.2 $\pm$0.9 km s$^{-1}$ \citep{wilson53}. Fig. \ref{figrv} displays the 
radial velocity points folded on the orbital period together with the best 
sine fit solution. New radial velocities are marked with open circles whereas 
solid circles indicate velocities from the old 1993-1997 database 
(C98). Our $K_2$ velocity is consistent  with C98 
within 1-$\sigma$ but not with E09 who finds $K_{2} = 79 \pm 3~ 
{\rm km s}^{-1}$. To ilustrate this point we also plot in
dashed-line style the best sinewave fit for $K_2=79$ km s$^{-1}$. This has   
$\chi^{2}_\nu=3$  for 60 degrees of freedom and hence 
it is far less significant than our $K_2$ value \citep{lampton76}. 

E09 have suggested that the difference between the two $K_2$ values 
could be caused by different levels of X-ray irradiation between the two
observing epochs. The detection of HeII $\lambda$4686 emission
from the companion by E09 certainly  indicates that the star is irradiated.
However, it remains to be seen whether irradiation is sufficient, not only to 
pump chromospheric HeII emission, but also to modify the surface distribution 
of the photospheric absorption lines. In principle, 
irradiation can quench absorption lines from the inner hemisphere of the donor
star, leading to an increase in the observed $K_2$ \citep{wadehorne88}. 
It might be possible that, by a chance coincidence, the E09 database was 
obtained during an episode of lower X-ray activity than our data and we have 
looked for this using RXTE/ASM (All Sky Monitor) data. Contemporaneous 
X-ray observations are 
available for the second half of the C98 campaign (since 1996, when RXTE was 
launched) our new data from 1999-2000 and the 
entire E09 database but comparable levels of X-ray activity are found, with the 
X-ray flux $F_{\rm X}$ oscillating between 35 and 50 ASM counts s$^{-1}$. Only 
one velocity point in C98, obtained on the night of 5 Aug 1996, was taken 
during a dip in the X-ray light curve of $\simeq$18 counts s$^{-1}$. 
Interestingly, its orbital phase is $\phi=0.63$, almost identical to that of 
another point obtained on 3 Aug 1997 ($\phi=0.67$), when  
$F_{\rm X}\simeq$ 50 counts s$^{-1}$ i.e. almost a factor 3 higher. Despite the 
difference in X-ray flux both velocities are consistent with our best 
orbital solution within 1$-\sigma$. And one should note that phase 0.65 is close 
to an orbital quadrature, when velocity distortions from a circular orbit should 
be largest \citep[e.g.][]{davey92}. This strongly suggests that the effects of 
X-ray irradiation in the radial velocity curve are unimportant. 

As a matter of fact, irradiation will distort the radial velocity curve from a
simple sine wave, introducing a fictitious eccentricity which
should be measurable. Therefore, we have also attempted to fit elliptical 
orbits to our database, following \cite{friend90}. Our best fit yields 
null eccentricity ($e=0.004 \pm 0.019$) and a larger reduced $\chi^2$ than a
simple circular solution, another indication that irradiation is negligible 
at this level. 
Furthermore, \cite{orosz99} find no evidence for excess light at phase 0.5 in
their optical light curves nor C98 observe any significant change of spectral
type with orbital phase. These two results also suggest that X-ray 
irradiation is not enough to explain the discrepant $K_2$ values. 
The lack of irradiation signatures is probably due to the fact that the donor 
star is hot and the orbital separation large \cite{orosz99}. The X-ray flux 
received by every surface element on the companion star is too small to produce 
any disturbance in the radial velocity curve or light curve, despite the near 
Eddington X-ray luminosity of Cyg X-2. 

Looking at the radial velocity curve of E09 (shown in fig. 4) we note a large 
scatter  in the velocity points near the phase 0.25 quadrature. Two datapoints 
seem to lie systematically lower than the rest by $\sim$20 km s$^{-1}$ and this 
is certaintly dragging the $K_2$ velocity to lower values. The authors admit 
that only one arc spectrum was obtained for most of the nights.  
Although sky lines were used to correct for instrumental offsets, the fact 
that the strongest sky line OIII $\lambda$5577 lies at the edge of their
spectral range may introduce some systematics in the offset correction.   
Therefore, we believe that the value reported by E09 might    
be affected by problems with the wavelength calibration which is a  
critical issue given their low spectral resolution of $\sim$160 km s$^{-1}$.  
In any case, new high resolution observations obtained at the two orbital 
quadratures are clearly required to further constrain $K_2$ and confirm our 
result. 

Since the donor star is filling its Roche lobe and synchronized, we can 
combine our updated $K_2$ and $V \sin i$ values to constrain the binary mass
ratio through 

\begin{equation}
V \sin i = K_{\rm 2}~(1 + q)~\frac{0.49~q^{2/3}}{0.6~q^{2/3}+\ln{(1+q^{1/3})}}
\end{equation}

\noindent
\citep{hornewade86} which leads to $q=0.34\pm0.02$. 
The revised mass function is 
thus $f(M)= M_{1} \sin^{3} i /(1+q)^{2} = P K_{2}^{3}/2 \pi {\rm G} = 0.66 \pm 
0.03$ and hence $M_{1} \sin^{3} i  = 1.19 \pm 0.06$ M$_{\odot}$. 
Assuming $i=62.5\pm 4^{\circ}$ \citep{orosz99} we find $M_{1}= 1.71 \pm 0.21$ 
M$_{\odot}$. 
The error budget is clearly dominated by the uncertainty in the
inclination angle and thus ellipsoidal model fits to new light curves are 
urgently needed to better constraint the stellar masses.

\section{Summary}

We have revisited the determination of the system parameters in Cyg X-2 with 21 
new high-resolution spectra obtained during 1999 and 2000. The new solution 
does not support the conclusions of E09 who claim for a
significantly lower value for the radial velocity semiamplitude of the donor
star. Instead, our results confirm previous determinations reported in 
C98.
The discrepancy cannot be explained by X-ray irradiation because 
the sinusoidal shape of the radial velocity curve is not 
disturbed. In particular, we find (i) no evidence for orbital  
eccentricity and (ii) no significant deviations between two velocity 
points at phase $\sim$ 0.65, when X-ray flux varies by a factor $\sim$3. 
Our refined orbital parameters are $P=9.84450 \pm 0.00019$ days, 
$K_2=86.5 \pm 1.2$ km s$^{-1}$ and $V \sin i = 33.7 \pm 0.9$ km s$^{-1}$, which
lead to $q=0.34\pm0.02$, $M_{1} \sin^{3} i  = 1.19 \pm 0.06$ M$_{\odot}$.  
Assuming $i=62.5 \pm 4^{\circ}$ from \cite{orosz99} leads to 
$M_{1}=1.71\pm 0.21$ M$_{\odot}$ and, therefore, the possibility that Cyg X-2 
harbours a neutron star more massive than canonical.

\section{Acknowledgments}

We thank the anonymous referee for helpful comments to the manuscript. 
MOLLY software developed by T. R. Marsh is gratefully acknowledged. Partly
funded by the Spanish MEC under the Consolider-Ingenio 2010 Program grant 
CSD2006-00070: first science with the GTC. J.C. acknowledges support from 
the Spanish Misitry of Sience and Technology through the project AYA2007-66887. 
J. I. G. H. acknowledges support 
from the EU contract MEXT-CT-2004-014265 (CIFIST). The INT is operated on the 
island of La Palma by the Isaac Newton Group in the Spanish Observatorio del 
Roque de Los Muchachos of the Instituto de Astrof\'\i{}sica de Canarias (IAC).

\begin{table*}
\centering
\caption[]{Log of the observations.}
\label{tabobs}
\scriptsize
\begin{tabular}{lccccc}
\hline
\hline
Date  & Instrument  & Wav. Range & Exp. time & Dispersion & Resolution \\
 & & $\lambda\lambda$ & (s) &  (\AA\ pix$^{-1}$) &(km s$^{-1}$)\\
\hline
25-26/07/1999 & UES  & 5300-9000  & 3200,3x(1800,2700,3600) & 0.11 & 10 \\
31/07/1999    & ISIS & 5885-6665  & 3x1800     & 0.45 & 34 \\
9/07/2000     & ISIS & 3545-4460  & 6x1800     & 0.45 & 44 \\
9/07/2000     & ISIS & 4350-5250  & 100, 1800  & 0.45 & 54 \\
\hline
\end{tabular}
\end{table*}

\end{document}